# Anisotropic nanoscale coherent polariton transport in CrSBr

P. Malik[1], D. Ko[1], V. Solovyeva[1], C. C. Palekar[2], I. Limame[2], S. Rodt[2], K. Mosina[3], Z. Sofer[3], A. Steinhoff[1], M. Esmann[1], S. Reitzenstein[2], B. Han[1], C. Gies[1], C. Schneider[1,*]

[1] *Institute of Physics, University of Oldenburg, D-26129 Oldenburg, Germany*
[2] *Institut fur Physik und Astronomie, Technische Universitat Berlin, D-10623 Berlin, Germany*
[3] *Department of Inorganic Chemistry, University of Chemistry and Technology Prague, Technicka 5, 166 28 Prague 6, Czech Republic*

* christian.schneider@uni-oldenburg.de

**In a combined experimental and theoretical study, we demonstrate anisotropic polariton transport on the nanoscale in the van der Waals antiferromagnet CrSBr. While effective cavity-polariton formation emerges via the self-hybridization of ultra-high oscillator strength excitons with a thin slab photonic mode, the absence of external mirrors facilitates spectroscopic investigation of these polaritons via cathodoluminescence (CL) on length scales determined by the electron wavelength. This direct access allows us to perform precise charting of the polariton landscape with nanometric resolution, and to probe polariton interference phenomena. The main finding of the work highlights that the coherent polariton transport follows the $C_{2v}$ symmetry of CrSBr, allowing exclusive transport along the crystallographic a-axis, while no coherent feature is found along the b-axis direction. Our work sets the foundation to use CL spectroscopy in cavity-polaritonics in more advanced landscapes, such as photonic crystals or optical lattices, and establishes the technique as a powerful tool to probe anisotropic expansion and relaxation phenomena on the nanoscale.**

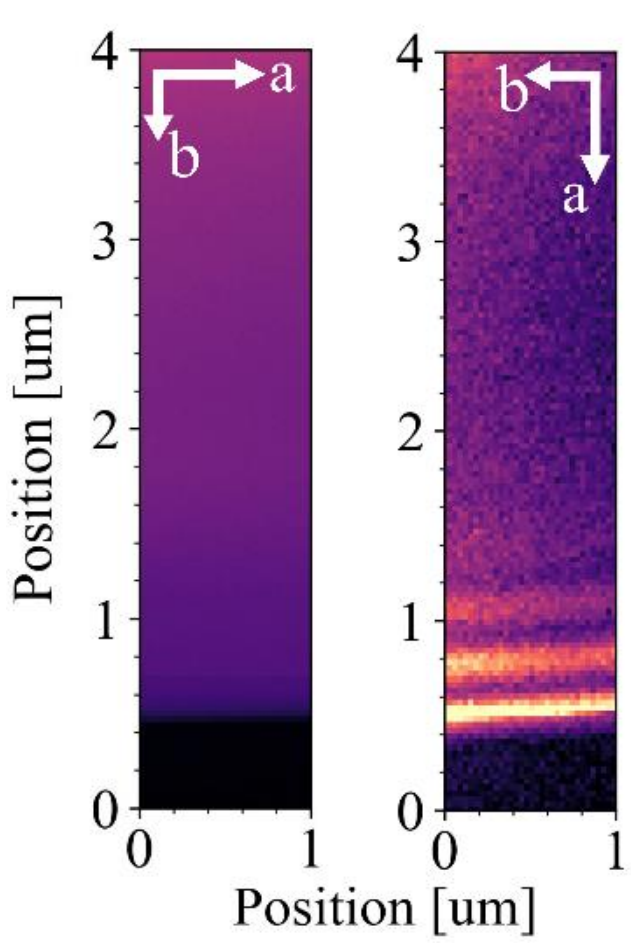

## INTRODUCTION

The magnetic van der Waals semiconductor Chromium Sulfide Bromide (CrSBr) has emerged as an exceptional platform to investigate strong light—matter interactions. This capability stems from its excitons with ultra-high oscillator strength [1]. Furthermore, the extraordinarily high refractive index of CrSBr allows slabs of sufficient thickness to inherently function as photonic Fabry-Pérot cavities, facilitating the formation of self-hybridized exciton-polaritons [2]. The orthorhombic crystal structure of CrSBr results in an anisotropic optical response, manifesting as strong birefringence [1]. Additionally, being an A-type antiferromagnet, the magnetic moments in CrSBr align in the same direction within one layer and alternate across layers [3]. The magnetic structure combined with the crystal anisotropy forces excitons to have quasi-1D nature [4]. As a result, excitons are linearly polarized along the crystallographic b-axis [1, 2, 4] which is also the easy magnetic axis. Secondly, the 1D nature manifests itself in the in-plane transport properties preferring the b-direction with an electric conductivity ratio ($\sigma_b/\sigma_a$) of the order of $10^5$ and the ratio of electron effective mass ($m_a/m_b$) reaching up to 50 [5, 6]. Interestingly, exciton transport has been shown to be anisotropic at 4 K, but becomes quasi-isotropic due to a magnon-drag effect at higher temperatures [7]. The self-hybridized polaritons, by virtue of the anisotropic electronic and optical landscape, are expected to inherit this stark in-plane anisotropy.

First works discussing the transport of self-hybridized polaritons in CrSBr have recently revealed anisotropic flow on the microscale [8]. However, far-field based photoluminescence (PL) techniques are unable to probe the nature of this transport on the nanoscale and any related emergent effects. Crucially, the self-hybridized nature of the polaritons in CrSBr eliminates the need for external mirrors for photonic confinement, permitting the use of localized excitation sources - like electrons - that entirely bypass the diffraction limit of light [9, 10].

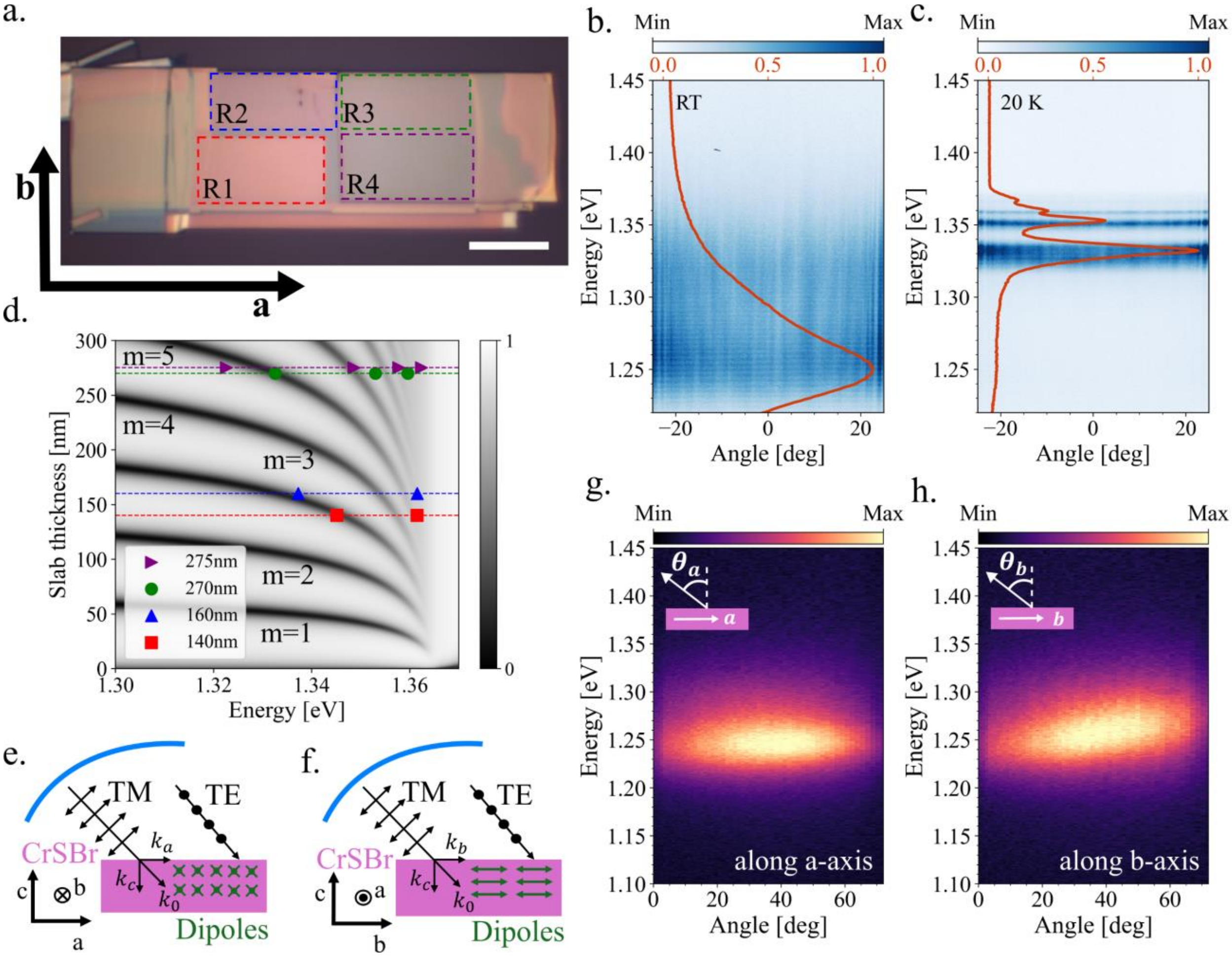

**Figure 1**. **a.** Microscope image of the 4-Region (R1-R4) CrSBr flake. Scale bar (white) is 10 $\mu m$. **b-c.** Angle-resolved photoluminescence (ARPL) overlaid with normalized CL spectra (red line) for R3 **b.** at room-temperature. **c.** at 20 K. **d.** Simulated reflectance at normal incidence for different slab thicknesses with the peaks extracted from CL spectra at 20 K for each of the four regions of the flake. **e-f.** Schematic diagram of the interaction of TE and TM polarized light with CrSBr slabs along **e.** a-axis and **f**. b-axis. The blue curve represents the axis of the parabolic mirror. **g-h.** Angle resolved cathodoluminescence (ARCL) in linear scaling for R3 **g.** along a-axis at room temperature. **h.** along b-axis at room temperature. Insets present a schematic of emission angle for each direction.

In this work, we use cathodoluminescence (CL) spectroscopy in conjunction with theoretical modeling to directly probe the landscape of the self-hybridized polaritons hosted by CrSBr slabs with thicknesses of 100 nm - 300 nm. In previous works, CL has been used to study nanoscale coherent expansion of waveguide polaritons in $WSe_2$ slabs [11]. However, it has not yet been used to chart the landscape of cavity polaritons, or to investigate the interplay between polariton expansion physics and material anisotropy. To address this gap, we probe the k-space anisotropy of the CL emission through angle-resolved measurements. More importantly, we perform hyperspectral CL maps on the edges parallel to the a and b crystal axes revealing a striking anisotropy in the coherent transport of the polaritons in the form of interference fringes.

## RESULTS & DISCUSSION

Our initial experiments focus on a CrSBr flake (Figure 1a) with four well-defined rectangular regions - R1, R2, R3, R4 - featuring approximate thicknesses of 140 nm, 160 nm, 270 nm and 275 nm, respectively. The flakes used in this study were all exfoliated using scotch tape and then transferred to $Si/SiO_2$ substrate using PDMS dry-stamping technique [12]. The thickness was measured using atomic force microscopy (AFM) (see methods section for details).

We begin by characterizing the flake comparing both PL and CL emission. Figure 1b-c presents the angle-resolved PL spectrum with an overlay of the CL spectrum (red) from region R3 at cryogenic and room temperatures, demonstrating excellent agreement between the emission spectra generated by optical and electron-beam excitation. This establishes the consistency among PL and CL techniques for exploring self-hybridised polaritons in CrSBr slabs. Furthermore, the extracted peak energies across all four regions are in line with our transfer-matrix simulations, see Figure 1d Using this plot, we can associate a mode number $(m)$ for the well resolved lowest emitting polariton. From this, we can estimate the Rabi splitting using the coupled oscillator model [2, 13]. For R3, the mode no. $(m)$ is 5 and cavity length $(L)$ is 270 nm resulting in a Rabi splitting $(\hbar\Omega_{\mathrm{R}})$ of 0.21 eV.

As the first step to address the interplay between the CrSBr material anisotropy and polariton propagation physics, we measured the angle-resolved cathodoluminescence (ARCL) spectrum at room temperature. This is enabled by inserting a slit to only select the emission reflected only from the axis of the parabolic mirror (PM) which is used to collect CL emission from flakes. The crystal axes were aligned with the PM axis, see Figure 1e-f. A beam energy(current) of 30keV(5nA) was used for excitation. The ARCL measurements reveal an anisotropy in the dispersion of the self-hybridized polaritons: The angle dependence of CL emission is significantly flatter when the PM is aligned with the a-axis compared to the b-axis, see Figure 1g-h. This can be understood by considering the biaxial birefringence of CrSBr: As illustrated in Figure 1e, when the PM is aligned with the a-axis, it probes the $k_b = 0$ plane in momentum space. In this configuration, TE-polarized cavity modes couple to the excitons. Since the angle between the polarization and the exciton dipoles remains constant, the modes experience a constant, extraordinarily high refractive index, $n_{TE} = n_b \sim 6.4$ [2], leading to the flat polariton dispersion seen in Figure 1g. Conversely, we probe the $k_a = 0$ plane when the PM is aligned with the b-axis. For this plane of incidence, only TM cavity modes couple with the excitons, and the angle between polarization vector and dipoles varies with $k_b$ (see Figure 1f) resulting in a varying refractive index ($n_{TM} = n_b n_c [n_b^2 \sin^2\theta +$ $n_c^2 \cos^2 \theta]^{-\frac{1}{2}}$, where $\theta$ is the angle of propagation inside the material [14]). For normal incidence, $n$ is the same as for TE in the previous configuration. For higher angles of incidence, $n$ decreases continuously. This reduction in the effective refractive index produces a lighter cavity photon and, consequently, a lighter cavity polariton [15] with a noticeably steeper dispersion at higher angles, perfectly in agreement with the data in Figure 1h.

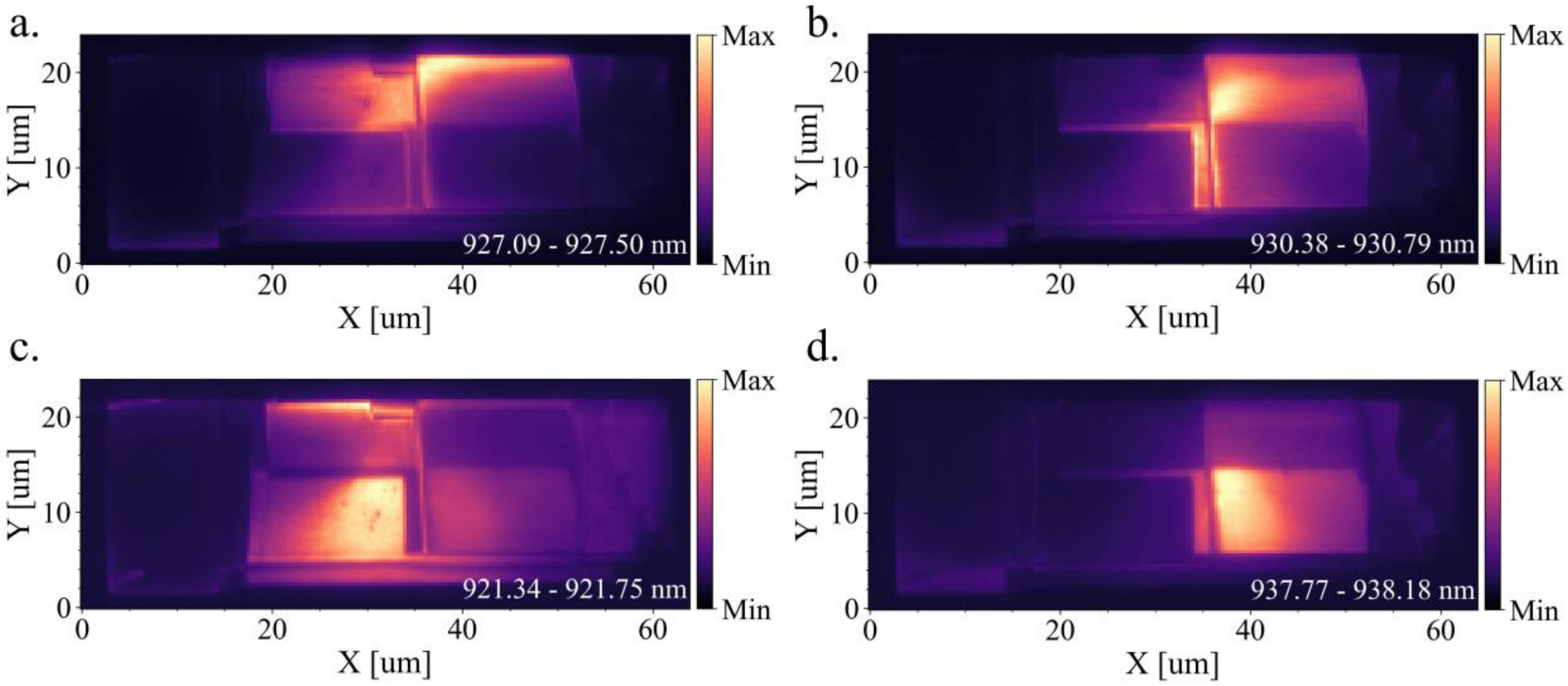

**Figure 2**. Hyperspectral CL raster map at 20K in linear scaling. Spatial resolution is 350nm **a-d.** are different slices of the maps at 927.3 nm (1.337 eV), 930.5 nm (1.332 eV), 921.5 nm (1.345 Ev), 938 nm (1.322 eV) respectively, normalised to the fully integrated CL image.

Mapping the polaritonic landscape at the nanoscale is a fundamental challenge for conventional optical methods, which are diffraction-limited and struggle to resolve sub-micron features. CL imaging overcomes this limitation. To demonstrate this, we perform hyperspectral CL raster imaging at cryogenic (20 K) and room temperature (~290 K). While room-temperature measurements suffer from significant broadening of the peaks, cryogenic temperatures drastically suppress this thermal broadening. The resulting sharpened linewidths of the CL emission become sufficiently narrow to completely mitigate spectral overlap. As theoretically anticipated by the simulated reflectance modes in Figure 1d, extracting spectral slices at specific wavelengths allows us to selectively image the emission in distinct regions of the flake shown in Figure 1a. Figures 2a-d present these targeted spectral slices, mapping the spatial distribution of the lowest-energy emitting polariton state for regions R2, R3, R1, and R4, respectively. Each slice is normalized against the fully integrated CL intensity map, revealing the spatial isolation of these polariton modes. This map was obtained with a spatial scanning resolution of 350 nm and a beam energy(current) of 20keV(5nA). It is interesting to note that, using CL, it was possible to distinctly resolve even the most intricate, sub-micron features, including the sharp terrace edges of the flake and the highly confined, thin, strip-like channel clearly visible in Figure 2b. Ultimately, this demonstrates the profound applicability of CL spectroscopy to spatially map polaritonic states over a microscale region, typical area of flakes and other photonic structures, while maintaining nanoscale resolution.

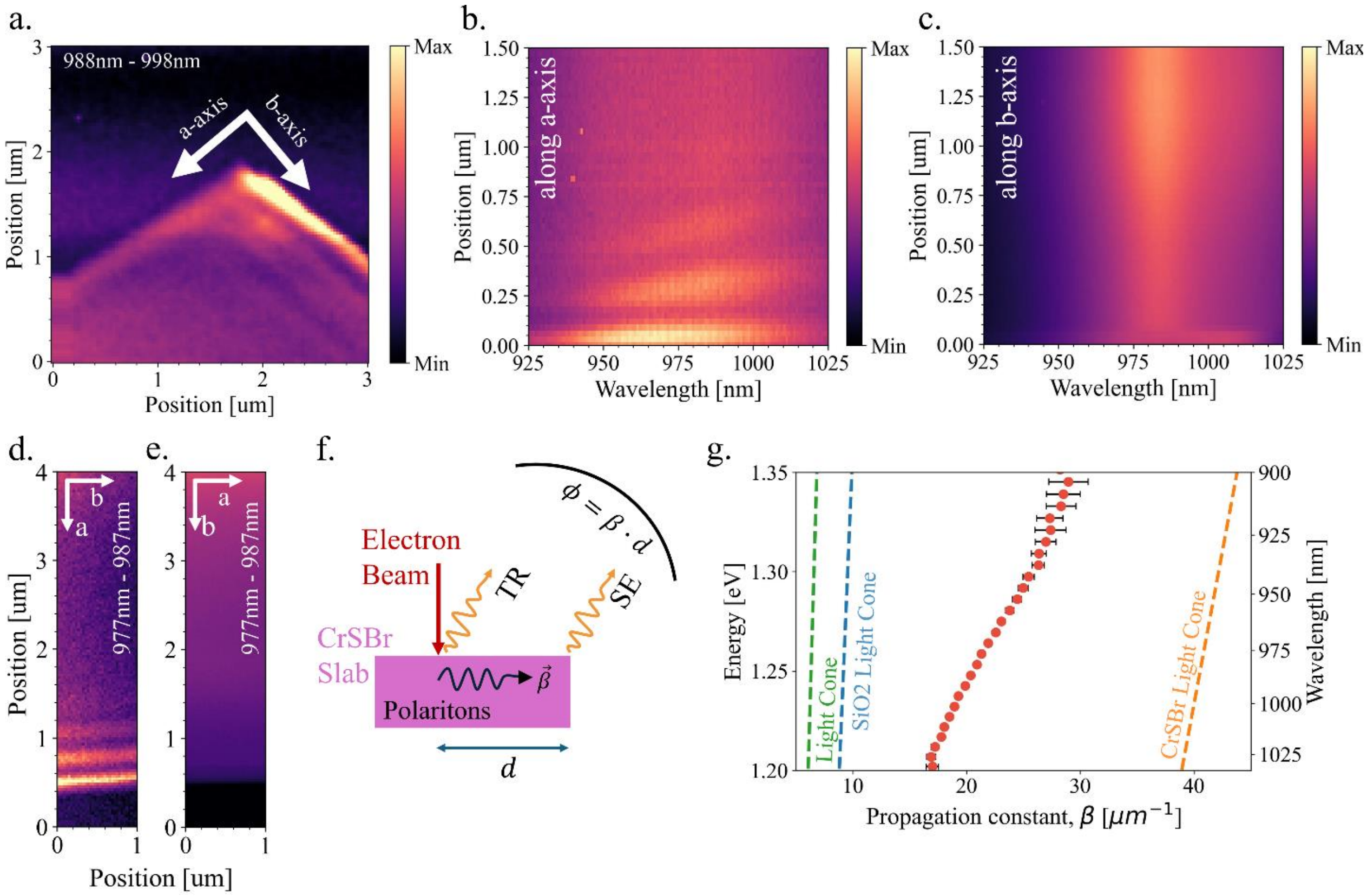

**Figure 3**. **a.** A slice of a hyperspectral CL map at the corner of a uniform thickness CrSBr flake at room-temperature in linear scaling **b-c.** Variation in the spatial CL emission with the emission wavelength and distance from the edge **b.** along the a-axis, fringes appear which are attributed to the far-field interference describe in panel f. **c.** along the b-axis, no fringes appear. **d-e.** Slice of hyperspectral CL map far from the corner **d.** along a-axis showing the fringes close to the edge **e.** along b-axis showing a flat emission **f.** Schematic diagram of the far-field interference scheme. **g.** Polariton E-k dispersion extracted from **b** overlaid with the vacuum and material light cones.

To explicitly probe the signatures of polariton transport anisotropy in CrSBr, we performed high-resolution hyperspectral CL raster imaging, focusing specifically on a right-angled corner of a flake with uniform thickness. These maps were obtained with a spatial resolution of 30 nm and a beam energy(current) of 20keV(5nA). Figures 3a presents a spectral slice that reveals clear interference fringes appearing along the edge parallel to the b-axis, which is strongly indicative for coherent polariton transport along the CrSBr a-axis. We assign these fringes to the far-field interference between the direct emission at the electron excitation spot called transition radiation (TR) [10], which we find to be anisotropic in the CrSBr crystal plane (see the following discussion of the theoretical results), and the scattered emission (SE) from waveguide polaritons propagating to the crystal edge and scattering to far-field (Figure 3f) [16]. The phase accumulated during the spatial transport of polaritons leads to a distinct modulation of the CL intensity, varying systematically with both the emission wavelength and the distance from the edge. While we observe strong intensity modulation for polaritons propagating parallel to the a-axis, indeed, no such modulation is present along the orthogonal b-axis. The suppression of propagation, revealed by the absence of fringes, can be directly attributed to the inherent anisotropy of the crystal. This effect is even clearer visualized in Figure 3b(c), where we present the CL intensity variation as a function of wavelength and position along a-axis(b-axis) away from the corner, and in Figure 3d(e) where we plot the spatial intensity variation for a spectral slice around 982nm corresponding to the Figure 3b(c).

We model the intensity modulations in terms of a phenomenological expression for the emission intensity $I$ derived from the two-point interference picture [9], which is governed by the distance $d$ from the edge,

$$I(\omega, d) = I_{bg}(\omega) + \left[A(\omega) \cdot \cos\left(\frac{\beta(\omega) \cdot d}{2}\right)\right]^2 \cdot \exp\left(-\frac{d^2}{l(\omega)^2}\right)$$

with the polariton coherence length $l(\omega)$, the polariton propagation constant $\beta(\omega)$, an arbitrary wavelength-dependent amplitude $A(\omega)$, and an incoherent background emission $I_{bg}(\omega)$

Figure 3g presents the dispersion relation (E-k plot) extracted from the experimental data in Figure 3b using this interference model. The group velocity of the polaritons propagating along a-axis is estimated from Figure 3g to be $\nu = (2.7 \pm 0.3)\ \mu m\ ps^{-1}$ at 975 nm. From the fit, we also extract a polariton coherence length of $l = (0.3 \pm 0.1)\ \mu m$ at 975 nm. We note that the directional preference of the polariton flow has also been observed as the long range polariton transport along the a-axis over a distance of 50 microns at a temperature of 4 K [8]. Here, we demonstrate this transport to be coherent up to a few hundreds of nanometres along the a-axis even at ambient conditions and heavily suppressed along the b-axis.

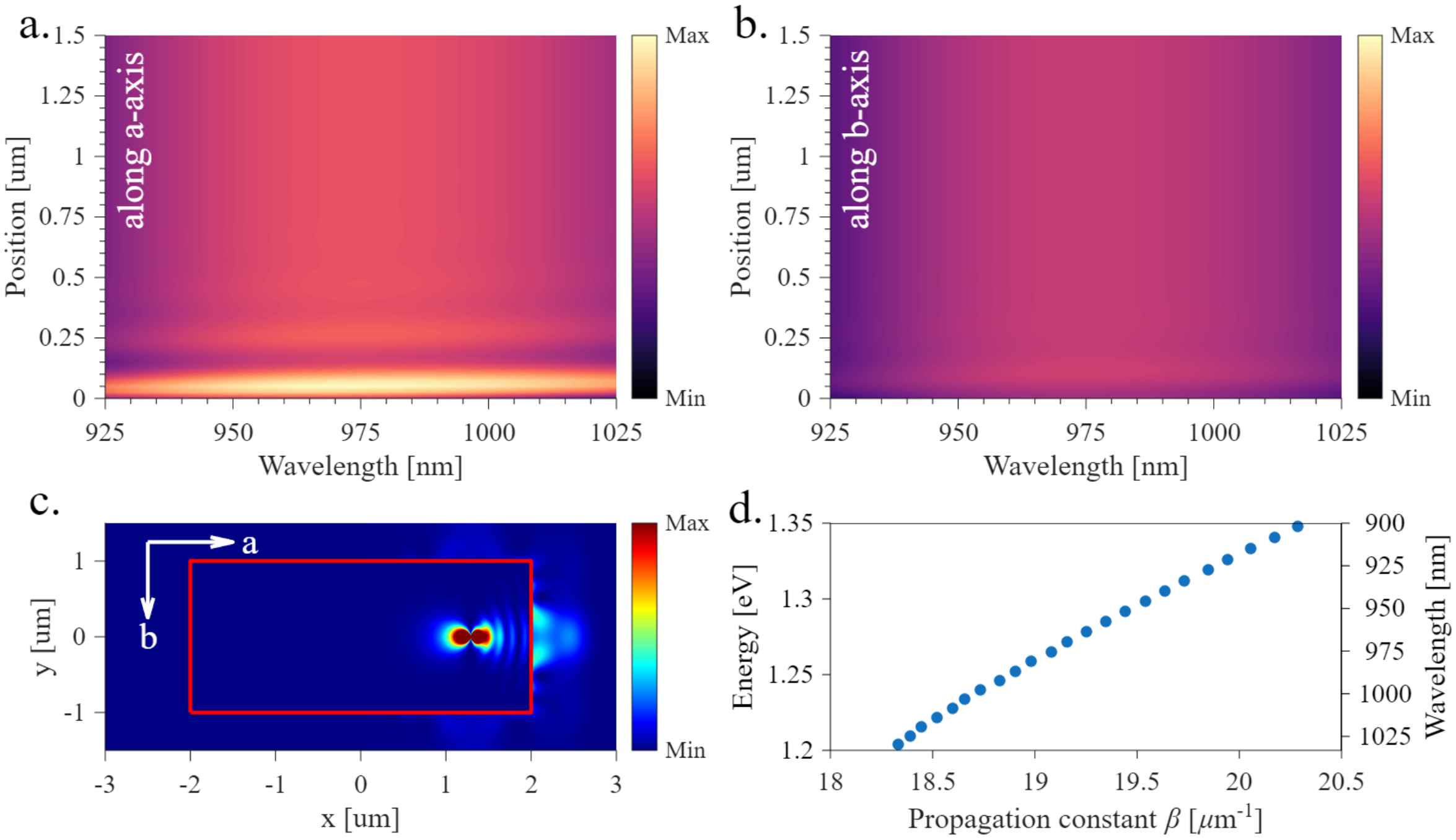

**Figure 4. a-b**. Simulated CL emission intensity profiles obtained from a numerical electrodynamical approach, mapping the wavelength-dependent real-space variations along the **a.** a-axis and **b.** b-axis of CrSBr, respectively in linear scaling. **c.** Electromagnetic power density profile simulated on the CrSBr sample, evaluated at a vertical position of z = 0 for a fixed wavelength. The red rectangle represents the edge of the CrSBr sample, which is surrounded by vacuum. **d.** Polariton propagation constant plotted against the corresponding photon energy (and corresponding wavelength). The dispersion relation is extracted via spatial Fourier transform analysis from the real-space power density data shown in (c), calculated across a range of different emission wavelengths.

To better understand the role that is played by the unique crystallographic anisotropy in the previously discussed results, we combine an electrodynamical approach used in [11] with a model fit of the CrSBr dielectric function [17] to simulate the electromagnetic field propagation in the sample after localized excitation. In a first step, we numerically solve a 2D Helmholtz equation to determine the vector potential $A$ induced by the electron beam, with the scalar potential $\phi$ following from a generalized Lorenz gauge condition. From the potentials, the associated electric and magnetic fields are calculated giving access to the z-component of the Poynting vector,

$$P_z = \frac{1}{2}\mathcal{R}e\left(E_x H_y^* - E_y H_x^*\right)$$

which determines the energy flow perpendicular to the sample. Key ingredient to this model is the transversal dielectric tensor in the optical spectral range, which encodes the excitonic response of the material to electromagnetic radiation leading to polariton propagation. The dielectric tensor of CrSBr is both anisotropic and strongly temperature dependent [17]. The optical response of the material along the b-axis is modelled using a complex dielectric function $\varepsilon_b(\omega)$ at room temperature (295K)

$$\varepsilon_b(\omega) = \varepsilon_\infty + \frac{A}{\omega_0^2 - \omega^2 - i\Gamma\omega} + \varepsilon^{TL}(\omega)$$

where $\varepsilon_\infty$ is the background dielectric constant and the second term represents the Lorentz oscillator for the primary (bulk) exciton, with the resonance frequency $\omega_0$, the oscillator strength $A$, and the spectral broadening $\Gamma$. The third term represents a sum over Tauc-Lorentz (TL) oscillators. In contrast to the b-axis, the dielectric function along the a-axis is frequency-independent and can be treated as a constant complex value. Further details of the model are provided in the SI.

Numerical results for the spatial emission versus wavelength and excitation position are presented in Figure 4, with panels (a) and (b) showing the variation of the excitation position along a- and b-axis, respectively. We find that the simulation successfully captures the optical anisotropy of CrSBr that was observed experimentally, see Figure 3b, c. When the beam is scanned along the a-axis, prominent interference fringes emerge. Conversely, scans along the b-axis show that interference features are strongly suppressed. The directional polariton propagation is determined by the pronounced anisotropy of the imaginary part of the dielectric function [8]. Because the imaginary component dictates optical absorption, it directly determines the propagation length of the excited polaritonic waves. For wave vectors along the a-axis, absorption is remarkably weak, allowing the waves to propagate freely over long distances, reach the crystal boundaries, and scatter into the far field. The interference between this edge-scattered light and the light emitted directly at the excitation spot forms the observed fringe pattern. In contrast, absorption is significantly higher for wave vectors along the b-axis. This strong damping prevents the waves from reaching the edges, thereby yielding a featureless propagation map. Besides explaining the observed anisotropic propagation, our electrodynamics description permits us to verify the intuitive picture of a two-point interference effect that was used above in the context of Figure 3f and g. The different sources of radiation are visualized in Figure 4c in terms of the two-dimensional power density distribution given by $P_z$. The red rectangle represents the edge of the CrSBr sample, where diffraction radiation emerges in the direction of polariton propagation. Also, the transition radiation emitted at the excitation spot is clearly visible, its elongated form being a signature of CrSBr in contrast to isotropic materials (a comparison is shown in the SI). Finally, by analyzing these real-space profiles, the polariton wavevector (propagation constant) is extracted via spatial Fourier transform and plotted against the photon energy (and corresponding wavelength) in Figure 4d to construct the dispersion relation. This plot naturally matches the E-k plot (Figure 3g) extracted from the experimental data in the previous section and validates the claim on the basis of which the phenomenological model captures the polaritonic propagation constant.

## CONCLUSION & OUTLOOK

In this study, we demonstrated and spatially mapped the anisotropic, coherent transport of self-hybridized exciton-polaritons in the van der Waals antiferromagnet CrSBr at the nanoscale. Additionally, we employed an electrodynamical approach to provide deeper insight into the interplay between polariton propagation and material anisotropy. This work also serves as a natural foundation to the exploration of hyperbolic polariton transport, a phenomenon which remains notoriously elusive to conventional far-field spectroscopy [17]. The high spatial resolution offered by CL makes it highly compelling to investigate more complex geometrical features that allow a similar access to an electron beam, such as polariton localization at the step-edges of CrSBr slabs that occur naturally during fabrication, or the behaviour of photonic crystal polaritons within transition metal dichalcogenide (TMDC) monolayers that inherit strong anisotropy from tailored cavity structures [18, 19]. Ultimately, this work establishes cathodoluminescence spectroscopy as a powerful tool for mapping polaritonic landscapes across advanced configurations, opening robust new pathway for investigating light-matter coupling at the nanoscale.

## METHODS

**Material:** CrSBr crystals were prepared by chemical vapor transport (CVT) in sealed quartz ampoules using chromium (99.99%, −60 mesh, Chemsavers, USA), bromine (99.9999%, Merck, Czech Republic) and sulfur (99.9999%, 2–6 mm, Wuhan Tuocai Technology Co. Ltd., China) as starting materials. The ampoules were filled with stoichiometric amounts of the starting components corresponding to 16 g of CrSBr, with an additional 2 at.% excess of sulfur and bromine. The ampoules were sealed under high vacuum while cooling the bromine-containing part with liquid nitrogen.

In the first step, the ampoules were gradually heated to 700 °C in a crucible furnace while keeping the upper part of the ampoule cold to ensure controlled reaction of bromine. After complete consumption of bromine, the ampoules were placed in a two-zone furnace for CVT growth. Initially, the growth zone was heated to 900 °C and the source zone to 750 °C for 2 days. Subsequently, the thermal gradient was reversed, and the source zone was heated to 900 °C while the growth zone was maintained at 800 °C for 14 days. After cooling to room temperature, the ampoules were opened in an argon-filled glovebox.

**AFM:** WiTec Alpha300RA in AC mode was used to characterize the thickness of the flakes.

**Photoluminescence:** The PL experiments were carried out using attoCube attodry800xs close cycle cryostat. We used a 633nm continuous-wave laser for excitation. A 50x Mitutoyo Plan Apo NIR infinitely corrected objective (NA=0.42) was used to collect the emission from sample, followed by a Fourier-space imaging lens configuration to resolve k-space emission. This emission spectrum is then measured using a Andor Shamrock 500i spectrometer with Andor iKon-M CCD.

**Cathodoluminescence:** For room-temperature CL experiments, we used a JEOL SEM equipped with a Delmic SPARC system and Andor Newton CCD, acceleration voltage 30kV, current 5nA. While for cryogenic experiments (20K), we used Raith eLine Plus electron beam lithography system equipped with a Helium-flow cryostat and a similar Delmic SPARC system configuration. Acceleration voltage 20kV, current 5nA.

**Details of the theoretical modeling are given in the SI.**

## ACKNOWLEDEGMENTS

*The work was funded by the European Union (ERC, Dual-Twist (101170213)). Views and opinions expressed are however those of the author(s) only and do not necessarily reflect those of the European Union or the European Research Council. Neither the European Union nor the granting authority can be held responsible for them."* and the Wissenschaftsraum ElLiKo (Volkswagenstiftung, Ministry of Science and Culture of Lower Saxony). Support by the German Research Foundation (DFG) with Project INST 1847/234-1 FUGG and by Berlin Quantum is acknowledged. The authors acknowledge the Deutsche Forschungsgemeinschaft (DFG, German Research Foundation) for funding the instrument JEOL IT800 scanning electron microscope project number INST 184/235-1 FUGG and for funding the WiTec Alpha300RA AFM INST 184/222-1 FUGG.

Z.S. was supported by ERC-CZ program (project LL2101) from Ministry of Education Youth and Sports (MEYS).